\documentclass[12pt]{article}
\usepackage{graphicx}
\usepackage{amssymb}
\usepackage{epstopdf}
\usepackage{hyperref}

\usepackage{amsmath,amssymb}
\usepackage{lmodern}
\usepackage{iftex}
\ifPDFTeX
  \usepackage[T1]{fontenc}
  \usepackage[utf8]{inputenc}
  \usepackage{textcomp} 
\else 
  \usepackage{unicode-math}
  \defaultfontfeatures{Scale=MatchLowercase}
  \defaultfontfeatures[\rmfamily]{Ligatures=TeX,Scale=1}
\fi
\IfFileExists{upquote.sty}{\usepackage{upquote}}{}
\IfFileExists{microtype.sty}{
  \usepackage[]{microtype}
  \UseMicrotypeSet[protrusion]{basicmath} 
}{}
\makeatletter
\@ifundefined{KOMAClassName}{
  \IfFileExists{parskip.sty}{%
    \usepackage{parskip}
  }{
    \setlength{\parindent}{0pt}
    \setlength{\parskip}{6pt plus 2pt minus 1pt}}
}{
  \KOMAoptions{parskip=half}}
\makeatother
\usepackage{xcolor}
\usepackage[margin=1in]{geometry}
\usepackage{color}
\usepackage{fancyvrb}

\DefineVerbatimEnvironment{Highlighting}{Verbatim}{commandchars=\\\{\}}
\usepackage{framed}
\definecolor{shadecolor}{RGB}{248,248,248}
\newenvironment{Shaded}{\begin{snugshade}}{\end{snugshade}}

\newcommand{\AttributeTok}[1]{\textcolor[rgb]{0.77,0.63,0.00}{#1}}

\newcommand{\ControlFlowTok}[1]{\textcolor[rgb]{0.13,0.29,0.53}{\textbf{#1}}}

\newcommand{\DecValTok}[1]{\textcolor[rgb]{0.00,0.00,0.81}{#1}}

\newcommand{\FunctionTok}[1]{\textcolor[rgb]{0.00,0.00,0.00}{#1}}

\newcommand{\NormalTok}[1]{#1}

\newcommand{\OtherTok}[1]{\textcolor[rgb]{0.56,0.35,0.01}{#1}}

\newcommand{\SpecialCharTok}[1]{\textcolor[rgb]{0.00,0.00,0.00}{#1}}

\newcommand{\StringTok}[1]{\textcolor[rgb]{0.31,0.60,0.02}{#1}}

\usepackage{graphicx}
\makeatletter
\def\maxwidth{\ifdim\Gin@nat@width>\linewidth\linewidth\else\Gin@nat@width\fi}
\def\maxheight{\ifdim\Gin@nat@height>\textheight\textheight\else\Gin@nat@height\fi}
\makeatother
\setkeys{Gin}{width=\maxwidth,height=\maxheight,keepaspectratio}
\makeatletter
\def\fps@figure{htbp}
\makeatother
\ifLuaTeX
  \usepackage{selnolig}  
\fi

\title{Comment on: Bell tests explained by classical optics without quantum entanglement}
\author{Richard D. Gill\\ \small Mathematical Institute, Leiden University}
\date{4 January, 2023\\{\small To appear in \emph{Physics Essays}}}                                           

\begin{document}
\maketitle

\begin{quote}
{\bf Abstract.} In a paper published in the journal {\em Physics Essays} in 2021, the author D.L. Mamas writes ``A polarized photon interacts with a polarizer through the component of the photon's electric field which is aligned with the polarizer. This component varies as the cosine of the angle through which the polarizer is rotated, explaining the cosine observed in Bell test data. Quantum mechanics is unnecessary and plays no role''. Mamas is right that according to this physical model, one will observe a negative cosine. However, the amplitude of the cosine curve is 50\%, not 100\%, and it consequently does not violate any Bell-CHSH inequality. Mamas' physical model is a classic local hidden variables model. The result is illustrated with a Monte Carlo simulation.

\bigskip
{\em Dans un article publié dans la revue {\em Physics Essays} en 2021, l'auteur D.L. Mamas écrit ``Un photon polarisé interagit avec un polariseur à travers la composante du champ électrique du photon qui est alignée avec le polariseur. Cette composante varie comme le cosinus de l'angle de rotation du polariseur, expliquant le cosinus observé dans les données de test de Bell. La mécanique quantique est inutile et ne joue aucun rôle''. Mamas a raison de dire que selon ce modèle physique, on observera un cosinus négatif. Cependant, l'amplitude de la courbe cosinus est de 50\%, et non de 100\%, et elle ne viole par conséquent aucune inégalité de Bell-CHSH. Le modèle physique de Mamas est un modèle classique à variables cachées locales. Le résultat est illustré par une simulation de Monte Carlo.}
\end{quote}

\section{Introduction}
In a paper published in {\em Physics Essays} in 2021, the author  D.L. Mamas writes ``A polarized photon interacts with a polarizer through the component of the photon’s electric field which is aligned with the polarizer. This component varies as the cosine of the angle through which the polarizer is rotated, explaining the cosine observed in Bell test data. Quantum mechanics is unnecessary and plays no role''\cite{mamas21}. In an earlier 2013 paper he writes ``the individual particles must be regarded as possessing intrinsic, fixed, and complementary quantum states even before an observer performs a measurement. There is no contradiction with quantum mechanics when calculated probabilities are viewed as purely statistical'' \cite{mamas13}.

He is right in supposing that his quasi-classical description would lead to a cosine shaped curve. Unfortunately, the correlation turns out to equal a factor $0.5$ times a cosine. Conventional quantum mechanics predicts (using the Born rule; there is no need to assume a non-local collapse of the wave function) a full cosine, the latter leading to violation of the Bell-CHSH inequalities. The reduced amplitude cosine curve however does satisfy Bell-CHSH inequalities, as it must do since Mamas' description of the physics of this experiment can be expressed as a local hidden variables model in Bell's sense.

This can of course be proven by working out the mathematical formulas using basic trigonometry and calculus, and this is done in the Appendix. We prefer to illustrate the result here, through a Monte Carlo simulation, using the open source and freely available R system.

\section{The simulation}

Imagine that two photons are generated at a source. Photon A has a polarization $\theta$ which is uniformly distributed between 0 and $\pi$ radians (180 degrees). Photon B has the orthogonal ``opposite'' polarization, $\theta + \pi/2$ radians. They head off to two polarisers which are set to orientations $\alpha$ and $\beta$ respectively, angles between  0 and $\pi$ radians. By the Malus law, photon A is then detected with probability $\cos^2(\alpha - \theta)$, photon B is detected with probability $\cos^2(\beta - \theta - \pi/2)$.

The following lines of code (Figure 1) implement this algorithm and generate the picture shown below (Figure 2). First of all, 100000 angles are drawn uniformly at random between 0 and $\pi$. Two further sets of the same number of random numbers, uniformly distributed between 0 and 1, are also created, in order to provide for the randomness at each polariser. We then fix Alice's polariser to direction 0, but generate a sequence of 1000 orientations for Bob, uniformly spaced between 0 and $\pi$. We create a vector of length 1000 to contain the 1000 correlations which we are going to empirically determine. For each of the 1000 orientations of Bob's polariser, we determine simultaneously for the 100000 photons at Alice's site, whether or not the photon passes the filter, and similarly for Bob. We can now average the products of the 100000 pairs of outcomes $\pm 1$, and save them in the vector of correlations. Finally, we plot the correlations.

\bigskip 

\newpage

\begin{Shaded}
\begin{Highlighting}[]
\NormalTok{thetas }\OtherTok{\textless{}{-}} \FunctionTok{runif}\NormalTok{(}\DecValTok{100000}\NormalTok{, }\DecValTok{0}\NormalTok{, pi)}
\NormalTok{xunifs }\OtherTok{\textless{}{-}} \FunctionTok{runif}\NormalTok{(}\DecValTok{100000}\NormalTok{)}
\NormalTok{yunifs }\OtherTok{\textless{}{-}} \FunctionTok{runif}\NormalTok{(}\DecValTok{100000}\NormalTok{)}
\NormalTok{alpha }\OtherTok{\textless{}{-}} \DecValTok{0}
\NormalTok{betas }\OtherTok{\textless{}{-}} \FunctionTok{seq}\NormalTok{(}\AttributeTok{from =} \DecValTok{0}\NormalTok{, }\AttributeTok{to =}\NormalTok{ pi, }\AttributeTok{length =} \DecValTok{1000}\NormalTok{)}
\NormalTok{corrs }\OtherTok{\textless{}{-}} \FunctionTok{rep}\NormalTok{(}\DecValTok{0}\NormalTok{, }\DecValTok{1000}\NormalTok{)}
\ControlFlowTok{for}\NormalTok{(i }\ControlFlowTok{in} \DecValTok{1}\SpecialCharTok{:}\DecValTok{1000}\NormalTok{)\{}
\NormalTok{  beta }\OtherTok{\textless{}{-}}\NormalTok{ betas[i]}
\NormalTok{  x }\OtherTok{\textless{}{-}} \FunctionTok{ifelse}\NormalTok{(xunifs }\SpecialCharTok{\textless{}} \FunctionTok{cos}\NormalTok{(thetas }\SpecialCharTok{{-}}\NormalTok{ alpha)}\SpecialCharTok{\^{}}\DecValTok{2}\NormalTok{, }\DecValTok{1}\NormalTok{, }\SpecialCharTok{{-}}\DecValTok{1}\NormalTok{)}
\NormalTok{  y }\OtherTok{\textless{}{-}} \FunctionTok{ifelse}\NormalTok{(yunifs }\SpecialCharTok{\textless{}} \FunctionTok{cos}\NormalTok{(thetas }\SpecialCharTok{+}\NormalTok{ pi}\SpecialCharTok{/}\DecValTok{2} \SpecialCharTok{{-}}\NormalTok{ beta)}\SpecialCharTok{\^{}}\DecValTok{2}\NormalTok{, }\DecValTok{1}\NormalTok{, }\SpecialCharTok{{-}}\DecValTok{1}\NormalTok{)}
\NormalTok{  corrs[i] }\OtherTok{\textless{}{-}} \FunctionTok{mean}\NormalTok{(x }\SpecialCharTok{*}\NormalTok{ y)}
\NormalTok{\}}
\FunctionTok{plot}\NormalTok{(betas, corrs, }\AttributeTok{type =} \StringTok{"l"}\NormalTok{, }\AttributeTok{ylim =} \FunctionTok{c}\NormalTok{(}\SpecialCharTok{{-}}\DecValTok{1}\NormalTok{, }\SpecialCharTok{+}\DecValTok{1}\NormalTok{))}
\FunctionTok{abline}\NormalTok{(}\AttributeTok{h =} \FunctionTok{c}\NormalTok{(}\SpecialCharTok{{-}}\DecValTok{1}\NormalTok{, }\DecValTok{0}\NormalTok{, }\DecValTok{1}\NormalTok{))}
\end{Highlighting}
\end{Shaded}
\noindent {\bf Figure 1}. R code for simulation of Mamas' physical model.

\bigskip

\bigskip

\bigskip

\bigskip

\includegraphics{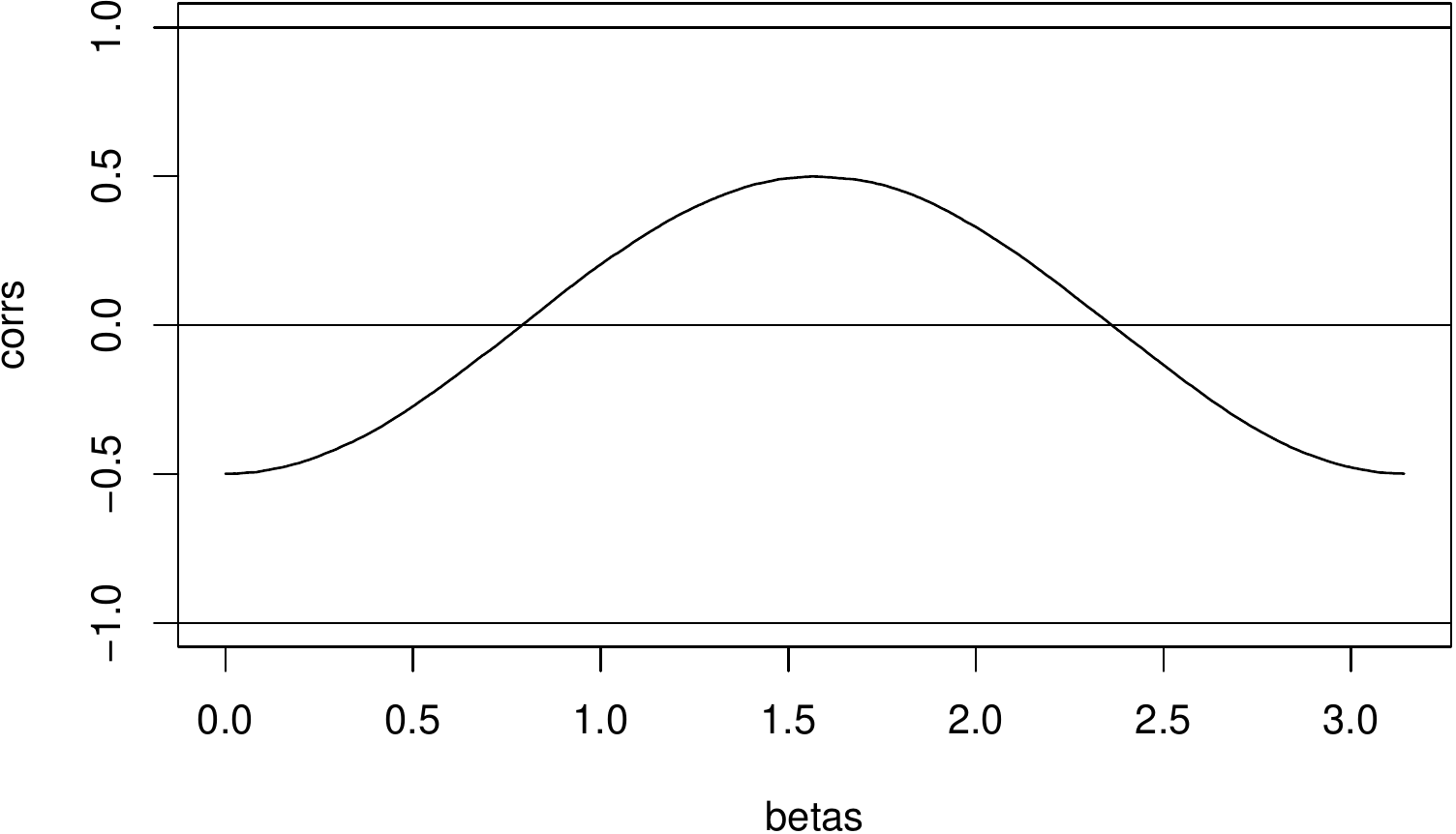}

\noindent {\bf Figure 2}. The correlation function according to Mamas' picture of the physics.

\newpage

\section{Concluding remarks}

Mamas' ideas are actually not novel. Already, Bohm and Aharanov discussed something like Mamas' physical model in their famous 1957 paper, transporting the EPR paradox to the case of two spin-half systems. That paper was the inspiration of John S. Bell's work, who turned the EPR, and the EPR-B, argument on its head by allowing Alice and Bob to measure differing observables, rather than only the same ones as one another. Al Kracklauer has published numerous papers exploring Mamas' basic idea in the 80's and 90's. It seems that Mamas does not understand the mathematical content of Bell's theorem, and is unaware of the huge literature Bell's work has engendered. Note that the standard QM computation of the EPR-B correlations uses Born's rule to get the joint probabilities of Alice and Bob's measurement outcomes, given their measurement settings, but does not require any assumption of non-local wave-function collapse. The metaphysical conclusions to be drawn from Bell's theorem are subject of endless debate to this day, but the mathematical core is irrefutable. Bell defines what he calls ``local hidden variables'', LHV, and shows that LHV implies Bell inequalities. A simple QM computation shows that the so-called EPR-B correlations found by Bohm and Aharanov do not satisfy those inequalities. Experiment shows that those correlations can however be observed in the real world; and in particular, even in the situation of well separated physical systems and under very rigorous experimental constraints (cf., the loophole-free Bell experiments of 2015 and 2016, \cite{hensen, giustina, shalm, rosenfeld} which led to the 2022 Nobel prize in physics). It is well known that under the assumption of LHV the largest amplitude of a negative cosine which could be observed in a typical Bell-type experiment in which settings of Alice and Bob are taken from the whole unit circle is about $70\%$, more precisely, $0.7071\dots = 1/\sqrt 2$. Mamas' assumptions define an LHV and give a negative cosine with amplitude $50\%$, smaller than $70\%$ and therefore in full agreement with Bell's theorem.

{
\raggedright

}

\section*{Appendix}

 I compute the probability that both photons will pass their respective polarization filters.
This is the average, as $\theta$ varies from from $0$ to $\pi$, of the probability that a photon
polarized in the direction $\theta$ passes a polarization filter set to the angle $\alpha$ and
simultaneously a photon
polarized in the direction $\theta + \pi$ passes a polarization filter set to the angle $\beta$,
thus equal to 
\begin{align*}
p(++)~&=~\frac1 \pi\int\limits_{\theta=0}^\pi \cos^2(\theta - \alpha)\cos^2(\theta + \pi/2 - \beta){\rm d}\theta\\
 &= ~ \frac1 {4 \pi} \int\limits_{\theta=0}^\pi(1 + \cos(2(\theta - \alpha)))(1 + \cos(2(\theta +\pi/2 - \beta))){\rm d}\theta\\
 &= ~ \frac14 +  \frac1 {4 \pi} \int\limits_{\theta=0}^\pi\cos(2(\theta - \alpha))\cos(2(\theta +\pi/2 - \beta)){\rm d}\theta\\
 &= ~ \frac14 +  \frac1 {8 \pi} \int\limits_{\theta=0}^\pi(\cos(4\theta - 2 (\alpha + \beta) + \pi) + \cos(- 2(\alpha-\beta) - \pi)){\rm d}\theta\\
 &= ~ \frac14 +  \frac1 {8 \pi} \int\limits_{\theta=0}^\pi \cos(2(\alpha-\beta) - \pi){\rm d}\theta\\
 &= ~ \frac14 -  \frac1 {8} \cos(2(\alpha-\beta)).
\end{align*}

\noindent That was the probability of ``+1, +1'', namely that both photons pass their detectors.
The marginal probability that each photon passes is one half. It follows that the probabilities of
``+1, –1'' and of ``–1, +1'' must both be $ \frac14 +  \frac1 {8} \cos(2(\alpha-\beta))$ and
the probability of ``–1, –1'' must be $ \frac14 -  \frac1 {8} \cos(2(\alpha-\beta))$, the same as
the probability of ``+1, +1''. The correlation (probability of equal outcomes minus probability
of opposite outcomes) is therefore $- \frac 12\cos(2(\alpha-\beta))$.

\end{document}